\documentclass[preprint,times]{elsarticle}

\usepackage{tikz}

\newcommand{\Alg}{{\cal A}}
\newcommand{\BAlg}{{\cal B}}
\newcommand{\proc}[1]{\mathrm{#1}}

\title{Work function algorithm can forget history without losing
 competitiveness}

\author{Livio Colussi}
\address{Department of Mathematics\\
University of Padova\\
via Trieste, 63\\
35121 Padova (Italy)
}

\begin{document}

\begin{abstract}
 The Work Function Algorithm (WFA) is the most effective deterministic
 on-line algorithm for the \(k\)-server problem. E.~Koutsoupias and
 C.~Papadimitriou in \cite{KouPap95A} proved that WFA is \( (2k-1)
 \)-competitive and it is conjectured that it is \( k
 \)-competitive. However the best known implementation of WFA requires
 time \( O(i^2) \) to process request \( r_i \) and this makes WFA
 impractical for long sequences of requests. The \( O(i^2) \) time is
 spent to compute the work function on the whole history \(
 r_1,\dots,r_{i} \) of past requests.
 
 In order to make constant the time to process a request, Rudec and
 Menger in \cite{RudecBM09} proposed to restrict the history to a
 moving window of fixed size. However WFA restricted to a moving
 window loses its competitiveness \cite{RudecM08}.
 
 Here we give a condition that allows WFA to forget the whole previous
 history and restart from scratch without losing
 competitiveness. Moreover for most of the metric spaces of practical interest
 (finite or bounded spaces) there is a constant bound on the length of
 the history before the condition is verified and this makes \( O(1)
 \) the time to process each request.
 
 The condition is first given for on-line algorithms in the more
 general framework of Metrical Task Systems (MTS)
 \cite{BoLiSa92,ChrLar98} ad then it is restricted to the \(k\)-server
 problem.
\end{abstract}


\maketitle

\section*{Introduction}  
  On-line algorithms process a sequence of inputs \(
  \rho=r_1,\dots,r_n \) on-line, i.e. they should process input \( r_i \) before the
  next input \( r_{i+1} \) is known. Sleator and Tarjan in
  \cite{SleTar85} introduced competitive analysis as a useful tool to
  evaluate on-line algorithms.
  
 Competitive analysis compares on-line algorithms \( \Alg \) against
 the best off-line algorithm \(\proc{Opt}\). \( \Alg \) is said \( \alpha
 \)-competitive if the cost \( C_{\Alg}(\rho) \) it pays to
 process an input sequence \( \rho \) is bounded by \(
 \alpha C_{\proc{Opt}}(\rho) \) where \(\proc{Opt}\) is an optimal
 off-line algorithm.
  
 Metrical Task Systems (MTS) was introduced by Borodin, Linial and
 Saks \cite{BoLiSa92} as a general framework for on-line
 algorithms. An MTS consists of a set \( S \) of \emph{states}, a cost \(
 d(s,t) \geq 0 \) to move from a state \( s \) to a state \( t \) and
 a cost \( c(r,s) \geq 0\) to process the input \( r \) in state
 \( s \).  We assume \( (S,d) \) is a metric space, i.e. \( d \) is
 symmetric, satisfies the triangle inequality and \( d(s,t)=0 \) iff
 \( s=t \).
 
 The metrical task system problem is as follows: given an initial
 state \( s_0 \) and a sequence \( \rho=r_1,\dots,r_n \) of inputs
 find a sequence \( \sigma=s_0,\dots,s_n \) of states that minimizes
 the total cost
 \[
  C(\rho,s_0) = \sum_{i=1}^n \left[d(s_{i-i},s_{i})+c(r_i,s_i)\right]
 \]
 
 The \( k \)-server problem is the problem of moving \( k \)
 servers around to service requests that appear on-line at points of a metric
 space. The goal is to minimize the total distance traveled by the
 servers.  The \( k \)-server problem  is defined by an initial
 configuration \( A_0 \) (the set of points where the
 servers are initially placed) and a sequence \( \rho = r_1,\dots,r_n
 \) of requests that appears at various points of the metric
 space. The \( k \) servers start in configuration \( A_0 \) and
 service requests \( r_1,\dots,r_n \) by moving through configurations
 \( A_1,\dots,A_n \) such that \( r_i \in A_i \) for all \(
 i=1,\dots,n \). The cost of a solution is the total distance traveled
 by the servers.
 
 The \( k \)-server problem can be seen as a particular MTS problem
 where the state set \( S \) is the set of server configurations \( A
 = a_1,\dots,a_k \), the distance \( d(A,B) \) is the minimal distance
 to move servers from configuration \( A \) to configuration \( B \)
 and \( c(r,A) = 0 \) iff a server in \( A \) is already in position
 \( r \) and \( c(r,A) = \infty \) otherwise.
 
 The Work Function Algorithm (WFA) is the most effective deterministic
 on-line algorithm with respect to the competitiveness. WFA was proved
 to be \( (2k-1) \)-competitive for the \( k \)-server problem
 \cite{KouPap95A} and \( (2|S|-1) \)-competitive for the general MTS problem \cite{BoLiSa92}.
 
 Despite WFA being very effective both in theory and in practice
 (an extensive testing of WFA can be found in \cite{RudecBM10}) it is seldom used in practice. The problem with WFA
 is that it needs to compute the work function for
 each input \( r_i \). However, work function depends on the
 whole previous history. This makes the computational complexity of the
 WFA
 prohibitive and ever-increasing.
 
 The fastest known implementation of WFA for the \( k \)-server
 problem is proposed in \cite{RudecBM09} and requires time \( O(i^2)
 \) to compute the work function at step \( i \) and time \(
 O(n^3) \) to process a sequence of \( n \) inputs.
 
 Many efforts have been made to find real-time on-line algorithms for
 the \( k \)-server problem, i.e. algorithms that require constant
 computational time to process each input. In \cite{BeiLar00}
 trackless algorithms are discussed and in
 \cite{BaumgartnerMH07,RudecM08} a WFA restricted to a moving window
 is proposed. However, in both cases, competitiveness is lost.
 
 Here we give a condition that allows any \( \alpha \)-competitive
 on-line algorithm \( \Alg \) for the MTS problem to discard the whole
 history and restart from scratch without losing
 competitiveness. More precisely we prove that for all \( \varepsilon
 > 0 \) we can discard history when the total cost paid becomes
 greater than \( 2\alpha(\alpha+\varepsilon)\Delta/\varepsilon \)
 where \( \Delta \) is an upper bound for state distance. The
 algorithm we obtain is \( (\alpha+\varepsilon) \)-competitive.
 
 Moreover, we show that, under very natural assumptions, there is a
 constant bound to the history length before the condition becomes true and this makes \( O(1) \) the time needed to
 process each input \( r_i \). For the MTS problem the assumption is
 that the state space is finite and there is a lower bound on the cost
 of processing an input. For the \( k \)-server problem the
 assumption is that there are lower and upper bounds for the length of
 a server motion.

\section{Discarding history while preserving competitiveness.}
 Let \( \Alg \) be any \( \alpha \)-competitive on-line algorithm for
 the MTS problem and let \( \rho \) be any
 sequence of inputs. Then \( C \leq \alpha W \) where \( C =
 C_{\Alg}(\rho) \) is the cost paid by the on-line algorithm to process the input sequence \( \rho \) starting from a given initial state \( s_0 \) and \( W = C_{\proc{Opt}}(\rho) \) is the cost paid by an optimal off-line algorithm \(\proc{Opt}\) to process the input sequence \( \rho \) starting from the same initial state \( s_0 \).   

 Assume the input sequence \( \rho \) is divided into \( m \geq 1 \) phases \( \phi_1, \dots \phi_m \) and let \( j_i \) the position of the last input \( r_{j_i} \) of phase \( \phi_i \). 
 
 Let \( \BAlg \) be an on-line algorithm that uses \( \Alg \) to process each phase separately, i.e. when the last request of a phase has been processed it restarts from scratch forgetting the whole previous history and using the final state of the previous phase as a new starting state.

 Let \( C_i = C_{\BAlg}(\phi_i) \) be the cost paid by algorithm \( \BAlg \) to process the \( i \)-th phase and \( Y_i = C_{\proc{Opt}}(\phi_i) \) the cost paid by an optimal off-line algorithm \(\proc{Opt}\) to process the same phase starting from the same initial state. Then \( C_i \leq \alpha Y_i \) (since \( \BAlg \) works as \( \Alg \) when processing a phase) and the cost \( C = C_{\BAlg}(\rho) \) paid to process the whole sequence satisfies the inequality
 \[
  C = \sum_{i=1}^m C_i \leq \alpha \sum_{i=1}^m Y_i
 \] 
 
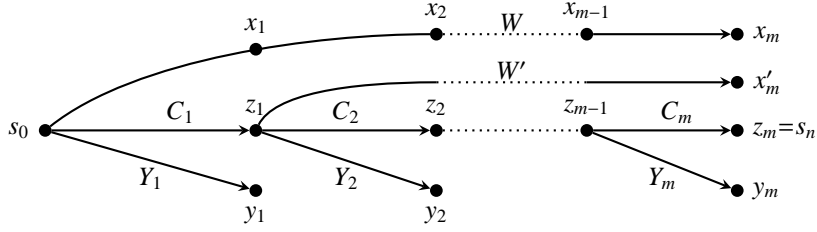
\begin{figure}%
\begin{center}
\begin{tikzpicture}
  [scale=0.8,
   vnero/.style ={circle,draw,thick,fill,
                  inner sep=0pt,minimum size=4pt}
  ]
  \node [vnero] (S0) at (0,0) 
        [label=left:$s_{0}$] {};
  \node [vnero] (S1)  at (3.5,0) 
        [label=above:$z_{1}$] {};
  \node [vnero] (S2)  at (6.5,0) 
        [label=above:$z_{2}$] {};
  \node [vnero] (Sn1) at (9,0) 
        [label=above:$z_{m-1}$] {};
  \node [vnero] (Sn) at (11.5,0) 
        [label=right:$z_{m}{=}s_{n}$] {};

  \draw[-stealth, thick] (S0) -- (S1);
  \node at (2.25,0.3) {$C_1$};
  \draw[-stealth, thick] (S1) -- (S2);
  \node at (5,0.3) {$C_2$};
  \draw[dotted, thick] (S2) -- (Sn1);
  \draw[-stealth, thick] (Sn1) -- (Sn);
  \node at (10.5,0.3) {$C_m$};

  \node [vnero] (X1)  at (3.5,1.35) 
        [label=above:$x_{1}$] {};
  \node [vnero] (X2)  at (6.5,1.6) 
        [label=above:$x_{2}$] {};
  \node [vnero] (Xn1)  at (9,1.6) 
        [label=above:$x_{m-1}$] {};
  \node [vnero] (Xn)  at (11.5,1.6) 
        [label=right:$x_{m}$] {};
  \draw[ thick] (S0)
   .. controls (1.5,1.2) and (4.5,1.6) .. (X2);
  \draw[dotted, thick] (X2) -- (Xn1);
  \draw[-stealth, thick] (Xn1) -- (Xn);
  \node at (7.75,1.8) {$W$};
    
  \node [vnero] (X1n)  at (11.5,0.8) 
        [label=right:$x'_{m}$] {};
  \draw[ thick] (S1)
   .. controls (3.8,0.6) and (5,0.8) .. (6.5,0.8);
  \draw[dotted, thick] (6.5,0.8) -- (9,0.8);
  \draw[-stealth, thick] (9,0.8) -- (X1n);
  \node at (7.75,1) {$W'$};

  \node [vnero] (Y1)  at (3.5,-1) 
        [label=below:$y_{1}$] {};
  \node [vnero] (Y2)  at (6.5,-1) 
        [label=below:$y_{2}$] {};
  \node [vnero] (Yn)  at (11.5,-1) 
        [label=right:$y_{m}$] {};
  \draw[-stealth, thick] (S0) -- (Y1);
  \node at (1.75,-0.8) {$Y_1$};
  \draw[-stealth, thick] (S1) -- (Y2);
  \node at (5,-0.8) {$Y_2$};
  \draw[-stealth, thick] (Sn1) -- (Yn);
  \node at (10.25,-0.8) {$Y_m$};

\end{tikzpicture}
\end{center}
 \caption{\label{BhAlg}
 \( C_1,C_2,\dots,C_m \) are the costs paid by \( \BAlg \) to process each phase separately restarting from scratch in states \( z_1,\dots,z_{m-1} \).
 \( Y_i \) is the cost paid by \(\proc{Opt}\) to process each phase \( \phi_i \) separately with initial states \( s_0,z_1,\dots,z_{m-1} \).
 \( W \) is the cost paid by \(\proc{Opt}\) to process the whole input sequence \( \rho \); \(x_1,x_2,\dots,x_m \) are the states where \(\proc{Opt}\) processes the last input of each phase. 
 \( W' \) is the cost paid by \(\proc{Opt}\) to process all together the last \( m-1 \) phases with initial state \( z_1 \). 
 }%
\end{figure}

 Figure \ref{BhAlg} shows the execution of \( \BAlg \) and, for each phase \( \phi_i \), the costs \( C_i \) and the final states \( z_i =s_{j_i} \). Below are the executions of \(\proc{Opt}\) on each phase \( \phi_i \) separately, the relative costs \( Y_i \) and final states \( y_{i} \). On top is the execution of \(\proc{Opt}\) on the whole input sequence \( \rho \), the cost \( W \), the final state \( x_m \) and the states \( x_{i} \) where the last input \( r_{j_i} \) of each phase \( \phi_i \) is processed. The execution of \(\proc{Opt}\) on the last \( m-1 \) phases together and the cost \( W' \) are also shown.

 Let \( W \) be split into the cost \( W_1 \) paid to process the first phase and \( W_2 = W-W_1 \).
 Then, by the minimality of \( Y_1 \)
 \[
  Y_1 \leq W_1-c(x_1,r_{j_1})+d(x_1,y_1) \leq W_1+d(x_1,y_1)
 \] 
 and by the minimality of \( W' \)
 \[
  W' \leq W_2+d(z_1,x_1)
 \]

 Let \( \varepsilon \) be any positive constant and assume the first phase satisfies the following condition:
 \begin{equation}\label{condition}
  C_1 \geq \frac{\alpha(\alpha+\varepsilon)(d(x_1,y_1)+d(z_1,x_1))}{\varepsilon}
 \end{equation}
 
 We can show, by induction on the number \( m \) of phases, that \( \BAlg \) is \( (\alpha+\varepsilon) \)-competitive. This is obviously true for \( m=1 \) since in this case \( \BAlg \) works exactly as the \( \alpha \)-competitive algorithm \( \Alg \).
 
 For \( m > 1 \) we have
 \(
  C_1 \leq \alpha Y_1
 \)
 and, by the inductive hypothesis
  \[
  \sum_{j=2}^m C_j \leq (\alpha+\varepsilon) W'
 \]
 
 Then
\begin{eqnarray*}
  W &=& W_1+W_2\\
    &\geq& Y_1+W'-d(x_1,y_1)-d(z_1,x_1)\\
    &\geq& \frac{C_1}{\alpha}+\frac{\sum_{j=2}^m C_j}{\alpha+\varepsilon}-d(x_1,y_1)-d(z_1,x_1)\\
   &=& \frac{C}{\alpha+\varepsilon}-\frac{C_1}{\alpha+\varepsilon}+\frac{C_1}{\alpha}-d(x_1,y_1)-d(z_1,x_1)\\
  &=& \frac{C}{\alpha+\varepsilon}+\frac{\varepsilon} {\alpha(\alpha+\varepsilon)}C_1-d(x_1,y_1)-d(z_1,x_1)\\
   &\geq& \frac{C}{\alpha+\varepsilon}+\frac{\varepsilon}{\alpha(\alpha+\varepsilon)}\frac{\alpha(\alpha+\varepsilon)(d(x_1,y_1)+d(z_1,x_1))}{\varepsilon}\\
    &&-d(x_1,y_1)-d(z_1,x_1)\\
  &=& \frac{C}{\alpha+\varepsilon}
\end{eqnarray*}
 Thus \( C\leq (\alpha+\varepsilon) W \) and \( \BAlg \) is \( (\alpha+\varepsilon) \)-competitive. 

 A problem in implementing algorithm \( \BAlg \) is that it should compare on-line the cost \( C \) paid so far to a bound \( \alpha(\alpha+\varepsilon)(d(x_1,y_1)+d(z_1,x_1))/\varepsilon \) that depends on state \( x_1 \) which is unknown at that point (since \( x_1 \) depends on future inputs). However, if the set of states is finite (as is usually assumed for MTS) there is an upper bound \( \Delta \) for state distance and we can instead test the condition
 \begin{equation}\label{condition1}
  C \geq \frac{\alpha(\alpha+\varepsilon)2\Delta}{\varepsilon}
 \end{equation} 
 that implies Condition \ref{condition}.

 We can also assume that there is a lower bound \( \delta \) for the cost \( c(r,s) \) to process an input \( r \) in a state \( s \).
 Then the cost \( C \) paid to process the first \( i \) inputs is at least \( i\delta \)  and condition \ref{condition1} is satisfied for 
\[
 i \geq \frac{\alpha(\alpha+\varepsilon)2\Delta}{\varepsilon\delta}
\]
 Thus there is a constant upper bound to the length of a phase and this makes \( O(1) \) the time to process each input \( r_i \).

 This result holds for all \( \alpha \)-competitive algorithms and so it also holds for the WFA that we known to be \( (2|S|-1) \)-competitive \cite{BoLiSa92}. Of course the last state of WFA and of the optimal off-line algorithm \(\proc{Opt}\) are the same. Referring to Figure \ref{BhAlg} we have \( y_i = z_i \) for all \( i \) and \( x_k=x'_k=z_k=y_k=s_n \). However this does not matter and does not help in Condition \ref{condition}.

\section{Bounded history WFA for the \( k \)-server problem.}
 The case of the WFA for the \( k \)-server problem is slightly more complicated. Costs \( c(r,s) \) are either 0 or \(\infty \) and states are configurations of servers i.e. sequences \( A = a_1,\dots,a_k \) of \( k \) points in the metric space \( V \) where the servers are moving. 

 Thus we cannot assume that the set of server configurations is finite and we cannot assume that there is a positive lower bound for the cost \( c(r,s) \). 

 In order to find an upper bound for distances \( d(x_1,y_1) \) and \( d(z_1,x_1) \) in condition \ref{condition} we can observe that server positions in configurations \( x_1 \), \( y_1 \) and \( z_1 \) can only belong to the \emph{set of interest}. The set of interest is the set of initial server positions and the position of requests processed so far.  
 
 Moreover in configurations \( x_1 \), \( y_1 \) and \( z_1 \) there is always a server in the position of the last served request. Then, if \( D \) is an upper bound for the distance between points in the set of interest we can bound \( d(x_1,y_1) + d(z_1,x_1) \) by \( 2(k-1)D \) thus obtaining condition
 \begin{equation}\label{condition2}
  C \geq \frac{2\alpha(\alpha+\varepsilon)(k-1)D}{\varepsilon}
 \end{equation}
 Notice that an upper bound \( D \) for the set of interest can be updated on-line in constant time if we choose a reference point (e.g. the position of server 1 in the initial configuration) and we take \( D \) equal twice the maximal distance of every other point of the set of interest to the reference point.
 
 We cannot say that condition \ref{condition2} will eventually be true since both \( C \) and \( D \) may grow. This is in accordance with \cite{RudecM08} where it is shown that WFA loses competitiveness when restricted to a fixed window.
 
 However, under normal circumstances \( D \) cannot grow indefinitely and there is an upper bound \( \Delta \) for it. This is always the case for finite or bounded spaces (e.g. finite graph). 
 
 The upper bound \( \Delta \) ensures condition \ref{condition2} to be eventually true but does not ensure that there is a constant upper bound for the history length before condition \ref{condition2} becomes true.
 To do so we need some more considerations.
 
 First of all, to prove competitiveness of an on-line algorithm for the \( k \) server problem we can only consider sequences \( \rho \) such that when request \( r_i \) appears no server is already in position \( r_i \) (removing such requests from the sequence \( \rho \) does not change the cost paid by the on-line algorithm while the cost paid by the optimal off-line algorithm can only increase).
 Thus we can avoid storing such requests in the history.
 
 Under normal situations we can also assume there is a lower bound \( \delta \) for the cost of an effective server move. 
 Then \( C \geq i\delta \) where \( i \) is the history length (without costless requests). Then condition \ref{condition2} is satisfied for 
\[
 i \geq \frac{2\alpha(\alpha+\varepsilon)(k-1)\Delta}{\varepsilon\delta}
\]
 and each request can be served in time \( O(1) \).

%}


\section*{Bibliography}

\begin{thebibliography}{9}
\bibitem{AcChNo00}
Dimitris Achlioptas, Marek Chrobak and John Noga.
\newblock Competitive Analysis of Randomized Paging Algorithms. 
\newblock \emph{Theoretical Computer Science - TCS}, 234, pp. 203--218, 2000.

\bibitem{BaumgartnerMH07}
Alfonzo Baumgartner, Robert Manger and Zeljko Hocenski.
\newblock Work function algorithm with a moving window for
		 solving the on-line k-server problem. 
\newblock \emph{Journal of Computing and Information Technology - CIT}, 15, pp. 325--330, 2007.

\bibitem{BeiLar00}
Wolfgang Bein and Lawrence L. Larmore.
\newblock Trackless online algorithms for the server problem. 
\newblock \emph{Inform. Process. Lett.}, 74:73--79, 2000.

\bibitem{BoLiSa92}
Allan Borodin, Nathan Linial and Michael Saks.
\newblock An optimal online algorithm for metrical task system. 
\newblock \emph{J.ACM}, 39:745--763, 1992.

\bibitem{ChrLar98}
Marek Chrobak and Lawrence L. Larmore.
\newblock Metrical task systems, the server problem,
			and the work function algorithm. 
\newblock In Amos Fiat and Gerhard J. Woeginger, editors. \emph{Online Algorithms: {The} State of the Art}, pages 74--94, Springer, 1998.

\bibitem{KouPap95A}
Elias Koutsoupias and Christos Papadimitriou.
\newblock On the $k$-server conjecture.
\newblock {\em J.ACM}, 42:971--983, 1995.

\bibitem{RudecBM09}
Tomislav Rudec, Alfonzo Baumgartner and Robert Manger.
\newblock A fast implementation of the optimal off-line algorithm for solving the k-server problem.
\newblock \emph{Mathematical Communications}, 14:119--134, 2009.

\bibitem{RudecBM10}
Tomislav Rudec, Alfonzo Baumgartner and Robert Manger.
\newblock Measuring true performance of the work function
		 algorithm for solving the on-line k-server problem. 
\newblock \emph{Journal of Computing and Information Technology - CIT}, 18:361--367, 2010.

\bibitem{RudecM08}
Tomislav Rudec and Robert Manger.
\newblock On the competitiveness of a modified work function
algorithm for solving the on-line k-server problem. 
\newblock In \emph{Proceedings of the ITI 2008 30th Int. Conf. on Information Technology Interfaces, June 23-26, 2008, Cavtat, Croatia}, pages 779--784, 2008.

\bibitem{SleTar85}
Daniel Sleator and Robert E. Tarjan.
\newblock Amortized efficiency of list update and paging rules. 
\newblock \emph{Commun. ACM}, 28:202--208, 1985.

\end{thebibliography}
\end{document}